\documentclass[4a,12pt]{article}
\usepackage{epsfig}

\begin{document}
\begin{center}
{\Large{\bf SO(10): a theory of fermion masses and mixings}}\\
[2ex]
{\large Goran Senjanovi\' c}\\
  ICTP, Trieste, Italy.\\
\end{center}
\begin{abstract}
SO(10) grand unified theory seems to have all the ingredients to be 
a complete unified theory of quarks and leptons. I review here its 
minimal, possibly realistic versions, both supersymmetric and not.
\end{abstract}

\section{Introduction}
   This talk I delivered in July last year in honor of Gustavo Branco
coming of age. Gustavo and I overlapped in graduate school
at City College of New York where we shared the same advisor
Rabi Mohapatra who also came of age last year. 
  City College in the seventies was a great place,
unique in a sense: you could do good physics and  at the same time 
teach local
kids from Harlem. And of course in the seventies
physics was really exciting, standard model, asymptotic 
freedom and all that, and one could think about L-R
symmetry, neutrino mass, strong CP and many other interesting
issues.

\section{SO(10): some generic features}
Grand unification, a theory of strong and electroweak interactions based 
on the single gauge group, implies two remarkable physical phenomena: 
proton decay and magnetic monopoles. More than 30 
years later neither 
has been found, a fact that seems to render the search of 
$\underline{\rm the}$ theory impossible. Fortunately, the accumulated 
information on fermion masses and mixings may provide the necessary 
clue. The minimal GUT group SU(5)  \cite{Georgi:1974sy} fails to unify the 
family of matter 
(the fermions) and in its minimal form predicts massless neutrinos. 
On the other hand, the minimal theory of matter unification, SO(10)
\cite{so10}, 
predicts massive neutrinos and through the seesaw mechanism 
\cite{seesaw} explains 
why neutrinos are so light. While SU(5) can be made to work with enough 
gymnastics, it by itself cannot connect quark and lepton mixings 
(for a recent study of ordinary SU(5) see e.g. 
\cite{Dorsner:2005ii}). The minimal predictive theory must be based 
on SO(10). 

Minimal or not, the SO(10) theory is certainly appealing on theoretical 
grounds. Besides unifying the family of fermions, and explaining the 
smallness of neutrino mass, it has left-right symmetry \cite{leftright} as a 
finite 
gauge transformations (broken spontaneously) in the form of generalized 
charge conjugation and a Pati-Salam SU(4)$_c$ symmetry which unifies 
quarks and leptons. The supersymmetric version has R-parity (matter 
parity) as a gauge symmetry \cite{rparity}, a part of the center 
$Z_4$ of SO(10). 
In some case (tree level see-saw) it can be shown \cite{lrsusy} that 
R-parity remains 
exact at all energies, surviving thus all the symmetry breaking. 
The lightest supersymmetric particle (LSP) is then stable, a perfect 
dark matter candidate. 

The fact that SO(10) by itself may naturally account for all fermion masses 
and connect small quarks and large lepton mixings is a strong argument
in its favor.
Here I offer a short updated review of the great effort in this 
direction. 
For more information and references see 
\cite{seesaw25,Mohapatra:2005zr}.

\section{SO(10) and only SO(10): can it be a complete theory 
(forgetting gravity)?}

The fermion families are 16-dimensional spinors of SO(10), and from

\begin{equation}
16_F\times 16_F=10_H+120_H+126_H
\end{equation}

\noindent
one has
 three possible Yukawa coupling matrices, 
enough to fit all the fermion masses and mixings. Now, if one wants a
predictive theory, one should stick to the minimal one. Better 
to say, one could use the information on masses and mixings to determine 
the theory, and thus make up for the 
absence of proton decay and monopole information. I want to 
describe and advocate this program  here. 

The point is simple. Ideally one could try a single Higgs, but a single 
set of Yukawas can be diagonalized and then all the fermion mass 
matrices would be simultaneously diagonal. Beside bad mass relations, 
this would imply no quark and no lepton mixings in the weak currents. 
The minimal theory must thus have two such Higgses (at least); in other 
words two Yukawa matrices. Compared to at least four in the standard 
model, it should be no surprise that such a theory is over constrained 
and predictive. In the next section I go through three such possible 
theories, one of which was studied at great length. The most 
interesting feature of these predictive theories is that they seem 
to determine the low energy (TeV) effective theory, i.e. decide 
whether or not one has a low energy supersymmetry, split 
supersymmetry \cite{split} or no supersymmetry at all. No need for 
philosophical arguments in favor or against supersymmetry.

This ambitious program of making  do without any new physics beyond the 
GUT 
itself is often praised. It is also often criticized 
for the same reason, namely for ignoring all the higher dimensional 
operators which would emerge from the physics above $M_{GUT}$. After 
all, such effects are in principle of the order $M_{GUT}/M_{Pl}\approx 
10^{-3}$, and thus potentially large compared to the first, and 
comparable to the second generation Yukawa couplings. However, 
such effects could be further suppressed by small dimensionless 
couplings. A nice example is the $d=5$ proton decay

\begin{equation}
O_{\Delta B\neq 0}=c QQQL/M_{Pl}\;,
\end{equation}

\noindent
which follows from the SO(10) invariant interaction

\begin{equation}
O_S=c 16_F^4/M_{Pl}\;.
\end{equation}

The proton longevity $\tau_p > 10^{33}$ yr implies $c < 10^{-6}$. 
Unless we play a texture game, with such small coefficients all 
other physical effects can be safely ignored. 

\section{The models}

We need to decide here which combination of $10_H$, $120_H$ and 
$\overline{126}_H$ is to be chosen (must be two out of three). 
We have the following possibilities:

(i) $\overline{126}_H+10_H$;

(ii) $120_H+10_H$;

(iii) $\overline{126}_H+120_H$;

(iv) $10_H+10_H$;

(v) $120_H+120_H$;

(vi) $\overline{126}_H+\overline{126}_H$;

We will need the Pati-Salam SU(2)$_L\times$SU(2)$_R\times$SU(4)$_C$ 
decomposition

\begin{eqnarray}
10_H&=&(2,2,1)+(1,1,6)\;,\\
\overline{126}_H&=&(2,2,15)+(1,3,10)+(3,1,\overline{10})+(1,1,6)\;,\nonumber\\
120_H&=&(2,2,15)+(2,2,1)+ (1,1,10) + (1,1,\overline{10}) +  (1,3,6) + (3,1,6)
\;,\nonumber
\end{eqnarray}

\noindent
and the fact

\begin{equation}
Y_{10}=Y_{10}^T\;,\;
Y_{126}=Y_{126}^T\;,\;
Y_{120}=-Y_{120}^T\;.
\end{equation}

With all three Higgs fields $10_H$, $120_H$ and $\overline{126}_H$, one
finds

\begin{eqnarray}
\label{matrices}
M_u&=&\langle 2,2,1\rangle_{10}^uY_{10}+
\langle 2,2,15\rangle_{126}^uY_{126}+
\left(\langle 2,2,1\rangle_{120}^u+
\langle 2,2,15\rangle_{120}^u\right)Y_{120}\;,\nonumber\\
M_d&=&\langle 2,2,1\rangle_{10}^dY_{10}+
\langle 2,2,15\rangle_{126}^dY_{126}+
\left(\langle 2,2,1\rangle_{120}^d+
\langle 2,2,15\rangle_{120}^d\right)Y_{120}\;,\nonumber\\
M_l&=&\langle 2,2,1\rangle_{10}^dY_{10}-
3\langle 2,2,15\rangle_{126}^dY_{126}+
\left(\langle 2,2,1\rangle_{120}^d-
3\langle 2,2,15\rangle_{120}^d\right)Y_{120}\;,\nonumber\\
M_D&=&\langle 2,2,1\rangle_{10}^uY_{10}-
3\langle 2,2,15\rangle_{126}^uY_{126}+
\left(\langle 2,2,1\rangle_{120}^u-
3\langle 2,2,15\rangle_{120}^u\right)Y_{120}\;,\nonumber\\
M_{\nu_R}&=&\langle 1,3,10\rangle Y_{126}\;,\nonumber\\
M_{\nu_L}^{II}&=&\langle 3,1,\overline{10}\rangle Y_{126}\;,
\end{eqnarray}

\noindent
where $M_u$, $M_d$, $M_l$, $M_D$, $M_{\nu_R}$, $M_{\nu_L}$ 
denote up quark, down quark, charged leptons, neutrino Dirac, 
right-handed neutrino and left-handed neutrino mass matrices 
respectively. The left-handed neutrino mass matrix $M_{\nu_L}^{II}$ 
is commonly called type II see-saw matrix 
\cite{Lazarides:1980nt,Mohapatra:1980yp}, because 
the 
induced 
vev for $\langle 3,1,\overline{10}\rangle$ is small: $\langle 3,1,\overline{10}\rangle \approx 
M_W^2/M_{GUT}$. Certain obvious features can be read-off from 
(\ref{matrices}):

a) 10 treats quarks and leptons on the same footing, since 
(2,2,1) is a $SU(4)_C$  singlet. This is ideal for the third generation
in the case of low energy supersymmetry.

b) $\overline{126}$ gives us the right-handed neutrino mass and the type I see-saw
(and also type II); and furthermore the Georgi-Jarlskog factor 
\cite{Georgi:1979df} 
$m_l=-3m_d$ since (2,2,15) is an adjoint of $SU(4)_C$.  This works 
well for the second generation.

c) In the absence of $\overline{126}$, neutrinos would only have a Dirac mass, 
and related to the charged fermion masses. This is cured through 
the introduction of $16_H$ needed to break $B-L$ 
anyway, since $16_H\times 16_H=126_H$ can simulate the direct 
presence of $\overline{126}_H$. 

It is easy to see that (iv) predicts $m_d=m_e$ and 
(vi) $3m_d=-m_l$ for all three generations. This is not 
correct. On the other hand the antisymmetry of 
$Y_{120}$ implies $m_1=0$ (the first generation mass) and 
$m_2=-m_3$, for the case (v). This is clearly wrong and so we are 
left only with (i)-(iii). Let us summarize next the possibilities, 
the exhaustive set of candidates for the minimal, realistic SO(10) GUT. 

\subsection{$\overline{126}_H+10_H$}

This model is emerging as the minimal supersymmetric 
SO(10) theory \cite{Aulakh:2003kg}. Although it has been around for 
more than two decades \cite{Aulakh:1982sw,Clark:1982ai}, 
it is only in the last three years that received the proper attention 
and there is now a dedicated effort in working out the predictions of 
the theory. It was noticed at the outset \cite{Lazarides:1980nt,Clark:1982ai}
that this minimal Higgs Yukawa structure could suffice. However it was
only a decade later that it emerged \cite{Babu:1992ia} that the theory 
can provide a deep connection between quark and lepton mixing angles,
and the information about the neutrino mass spectrum.

This program was boosted with the observation on a connection 
between a large atmospheric neutrino mixing angle and $b-\tau$ unification 
\cite{Bajc:2002iw,Bajc:2004fj,Bajc:2001ef} in the case 
of type II see-saw. This is easily seen from 
(\ref{matrices})\cite{Brahmachari:1997cq}

\begin{equation}
M_d-M_l\propto Y_{126}\propto M_{\nu_L}^{II}\;.
\end{equation}

Let's illustrate what happens for the case of $2^{nd}-3^{rd}$ 
generations. 

In the basis for diagonal charged leptons, and for the small down 
quark mixings (see \cite{Bajc:2002iw}),

\begin{eqnarray}
\label{mnu}
M_\nu\propto\pmatrix{
  m_\mu-m_s
& \epsilon
\cr
  \epsilon
& m_\tau-m_b
\cr}\;.
\end{eqnarray}

Clearly, a large atmospheric angle requires $m_b\approx m_\tau$.
 This illustrates nicely how a spontaneously 
broken quark-lepton asymmetry naturally accounts for the small 
quark and large lepton mixing angles. This is often claimed in 
the literature to be a mystery for no good reason at all since 
in the SM there is no connection between quark and lepton properties 
and certainly no reason to have the same or similar mixings. After 
all, the neutrino mass ratios are completely different from the 
charged fermion ones (much less hierarchical); if the mixings are 
related to mass ratios as hoped, it is more natural to have the 
mixings rather different. 

In any case, we see here that the unified theory of quarks and 
leptons does precisely that: connects large lepton with small 
quark mixings. In the following discussion we will offer two 
more examples which account for the same phenomena, but with 
different predictions for the masses. 

The simplified two-generation analysis above must be performed 
numerically for the case of three generations. Both type I and 
type II scenarios seem to work generically and what emerges is 
the prediction of hierarchical neutrino masses and an appreciable 
leptonic 13 mixing angle: $\left|U_{e3}\right|\ge 0.1$  \cite{Goh:2003sy,Bertolini:2005qb,Babu:2005ia}  . 
We have an example here of a predictive theory, good enough to be 
ruled out. It is remarkable that such a constrained Yukawa sector can
account correctly for all the fermion masses and mixings. 

However, 
if one restricts himself to the case of the minimal GUT Higgs sector 
too, type II seems to run into trouble \cite{Aulakh:2005sq,charan,Aulakh:2005bd,lagra,b2d} and 
possibly type I too \cite{b2d}. The problem is the compatibility of
mass and mixings fittings with unification constraints,  a study facilitated 
by the detailed computation of the full particle spectrum and couplings
\cite{Aulakh:2002zr,Fukuyama:2004ps,Bajc:2004xe}.  This was confirmed
recently for the general case, with the same tension between
the fermion mass fits and the gauge coupling unification or/and
proton decay \cite{Bertolini:2006pe}.  Thus this particular version
of the minimal SO(10), often coined the minimal supersymmetric 
SO(10) theory, seems to be in trouble. 

   However, all this is done by ignoring the effects of soft supersymmetry breaking. 
Clearly, such terms have a non negligible impact on the first and possibly even the
second generation fermion masses. Until they are included, one cannot proclaim
the above theory wrong; the trouble is that for generic soft terms the theory is not
predictive any more. One could attempt a simpler possibility of flavor blind soft terms
at high scale as in gauge mediation models, but even more interesting is to take the
theory seriously even as a source of supersymmetry breaking and its transmission to
the light sector, along the lines \cite{Bajc:2006ka}. This rather involved project is being 
planned.

\subsection{$\overline{126}_H+120_H$}

Another interesting possibility. It is in contradiction with low energy 
supersymmetry, since it leads to a prediction $m_\tau\approx -3 m_b$ 
at $M_{GUT}$, far from the $m_b\approx m_\tau$ of the MSSM. It could 
in principle work in the ordinary, non-supersymmetric SO(10), where 
$2 \left| m_b\right|\approx |m_\tau |$ at $M_{GUT}$. Namely, there 
are $m_s/m_b$ ($m_\mu / m_\tau$) effects that should be included, 
and also in ordinary SO(10), $M_R$, the SU(2)$_R$ breaking scale 
is around $10^{13}$ GeV or so, and thus the effects of right-handed 
neutrinos between $M_R$ and $M_{GUT}$ may be appreciable \cite{Vissani:1994fy}.
 For this reason we have recently studied this model in the context of the ordinary 
SO(10) \cite{Bajc:2005zf}.

Again, it is rather useful to get an analytical insight by focusing 
on the $2^{nd}$ and $3^{rd}$ generations only. 
Our findings are the following.

(a) 

\begin{equation}
\label{dmsm}
\frac{m_3^2-m_2^2}{m_3^2+m_2^2}=
\frac{\cos2\theta_A}{1-\sin^22\theta_A/2} + {\cal O}(|\epsilon|)
\end{equation}  

(b)

\begin{equation}
|V_{cb}|= |\ {\rm Re}\xi -
i\cos2\theta_A\ {\rm Im}\xi  |+ {\cal O}(|\epsilon^2|)
\end{equation} 

\noindent
where $\xi=\cos2\theta_A\  (\epsilon_d-\epsilon_u)$ and $\epsilon_i$ are 
the ratios between the relevant $2^{nd}$ and $3^{rd}$ generation 
masses.

(c)

\begin{equation}
\frac{m_\tau}{m_b}=3 + 3 \sin 2\theta_A\ {\rm Re}[\epsilon_e-\epsilon_d]
+{\cal O}(|\epsilon^2|)
\end{equation}

at $M_{GUT}$. The prediction (a) is rather interesting, connecting 
the neutrino degeneracy with the largeness of the atmospheric mixing 
(which cannot obviously be maximal). The prediction (b) needs appreciable 
corrections in order to work. It connects nicely small quark and large
lepton mixing, but ironically 
the  quark mixing turns out too small.
 
Also, (c) as we commented before needs 
corrections (to be computed), since in the SM $2 |m_b|\approx |m_\tau |$. 
A careful three generation numerical study is needed before one can know 
whether or not this theory works.

\subsection{$10_H+120_H$}

\begin{figure}[h]
\includegraphics[width=0.70\linewidth]{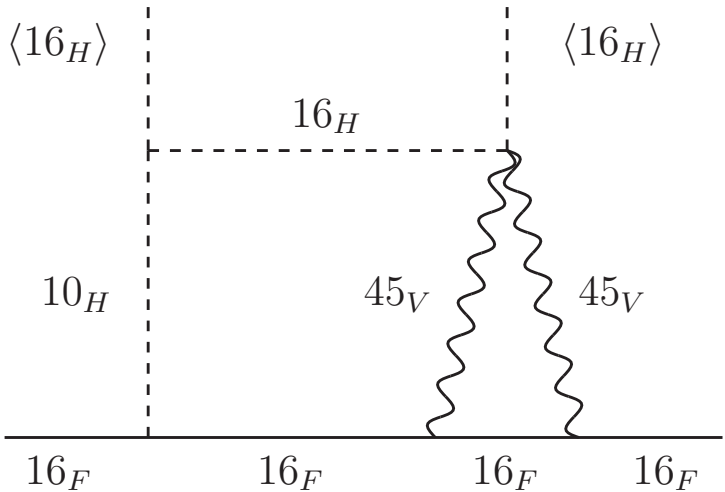}
\caption{\label{fig1} A contribution to the radiatively 
generated fermion mass.}
\end{figure}

In this case one must use $16_H$ instead of $126_H$ 
in order to break $B-L$ and give $\nu_R$ a mass. This allows for a 
radiative see-saw \cite{Witten:1979nr} at the two-loop level (Fig. \ref{fig1}), 
where 
one utilizes the gauge bosons in $45_V$ and $10_H$ in order to 
generate $126_H$ effectively. This beautiful mechanism requires 
supersymmetry to be strongly broken due to the non-renormalization 
of the superpotential. 

One can estimate

\begin{equation}
m_{\nu_R}\approx\left(\frac{\alpha}{\pi}\right)^2Y_{10}\frac{M_R^2}{M_{GUT}}
\frac{\tilde{m}}{M_{GUT}}\;,
\end{equation}

\noindent
where $M_R$ is the SU(2)$_R$ breaking scale, i.e. $\langle 16_H \rangle$ 
and $\tilde{m}$ the effective supersymmetry breaking scale in the visible sector (if 
$\tilde{m} > M_{GUT}$, then of course one would have $M_{GUT}/\tilde{m}$ 
suppression). 

In order to get $m_{\nu_R}$ as large as possible, $\tilde{m}$ should be 
as large as possible, which at first glance prefers no supersymmetry 
at all. However, then $M_R\ll M_{GUT}$ in this case, since the single 
step unification does not work in the SM. Furthermore, one gets into 
trouble with $b-\tau$ unification.

For a study of this model
see   \cite{Bajc:2004hr,Bajc:2005aq,Bajc:2006pa}, 
but under reasonable conditions one obtains the same 
predictions (a) and (b) as in (ii), with (c) $m_b\approx m_\tau$ at 
$M_{GUT}$. The last prediction favors a split susy scenario with 
$\tilde{m}\approx 10^9-10^{12}$ GeV (see Ref. \cite{Giudice:2004tc}
for a discussion of $m_b\approx m_\tau$ in split susy).

Recently, the supersymmetric version of this Yukawa sector was studied in the 
context of charged fermions \cite{Aulakh:2005mw,Lavoura:2006dv}. Notice that the 
results in these papers do not apply to the theory we are discussing here, since in 
order to work in this case the situation favors strongly split supersymmetry. Actually,
these works use the $126_H$ for neutrinos, but decouple it from the charged fermions.
This I find contrary to the main point  of SO(10), i.e. the connection 
between Dirac and Majorana Yukawas or the connection between charged fermions 
and neutrino masses. To me, it makes sense in the Pati-Salam theory where there is
no connection between the two, but SO(10) is there in order to cure this.

 \section{Summary and outlook}

I have tried in this short review to make a strong case for grand unification as a 
complete phenomenological theory (gravity ignored). The lack of observation of 
proton decay and magnetic monopoles made it for a long time basically impossible to 
define the minimal complete GUT. The hope is emerging now though through the 
information on fermion masses and mixings. Small non-vanishing neutrino masses point 
strongly towards SO(10), and SO(10) alone, without any new additional physics, 
offers predictive models. The minimal versions of the theory are good enough to be 
ruled out and the interesting model with low energy supersymmetry seems not to 
survive the fermion mass fitting and the unification constraints.  Here though one 
must include the effects of supersymmetry breaking before claiming this model dead. 
The proper approach is to study this within the model which would provide a 
predictive model of soft terms. We plan to address this tedious project in near 
future.

  What about giving up minimality?  This was discussed off and on in the past, 
and recently again a case was made for a superymmetric model with a full Yukawa 
sector in \cite{Aulakh:2005mw, Grimus:2006bb,Aulakh:2006vj}. Not surprisingly these 
extended versions seems to pass the tests (they also seem to work in rather 
restricted portions of the parameter space); after all the minimal version almost 
made it and failed only when the tough constraints of the minimality of even the GUT 
Higgs sector were included. However, one should complete in my opinion the study of 
the other two equally minimal versions, the non-supersymmetric and the split 
supersymmetric ones, before giving up on the minimal theory. If finally one of the 
three possibilities I discussed survives, it will be important to compute all the 
proton decay rates.

It is interesting to note that the search for proton decay led to the discovery of 
atmospheric neutrino oscillations. Could it be that similarly on the theoretical 
side grand unification will turn out to be the theory of neutrino masses and mixings 
before being the theory of proton decay and magnetic monopoles? In any case, this 
program will succeed only if it manages to connect these phenomena in a predictive 
manner.

\section{Added note}

As I was preparing this manuscript for the net, two new papers appeared devoted
to the in-depth study of the non-minimal model above, supplemented with the
spontaneous CP violation \cite{Aulakh:2006hs}. This constraints the theory
substantially and appears to be worth exploring.

Also, meanwhile a rather predictive theory emerged \cite{Bajc:2006ia} based on a 
simple extension of the minimal SU(5). It utilizes the so-called type III seesaw 
\cite{Foot:1988aq}, and predicts light fermionic weak triplet with mass below TeV. 
This offers a remarkable possibility of seesaw possibly directly probed at LHC, 
since the decay of the triplet into charged leptons goes through its neutrino Yukawa 
couplings. Although this theory may lack the beauty and the depth of SO(10), its 
predictivity and the relevance for LHC makes a strong case in its favor.

\section{Acknowledgements}

 I thank Gui Rebelo for a very nice Gustavo Fest, an excellent meeting
in a beautiful setting. The work described here was done in collaboration
with my friends Charan Aulakh, Borut Bajc, Alejandra Melfo and Francesco 
Vissani (AMIGO collaboration) to whom I am deeply grateful.

\end{document}